\documentclass[aps,11pt,nofootinbib,amsfonts,amssymb,showpacs]{revtex4} 
\usepackage{epsfig}
\usepackage{graphics}
\usepackage{color}

\begin{document}

%%%%%%%%%%%%%%%%%%%%%%%
\newcommand\new[1]{\ensuremath{\blacktriangleright}#1\ensuremath{\blacktriangleleft}}
%%%%%%%%%%%%%%%%%%%%%%%

%%%%%%%%%%%%%%%%%%%%%%%%%%%%%%%%%%%
\title{Geometry and stability of dynamical systems}

\author{Raffaele Punzi}
\email{raffaele.punzi@desy.de}
\affiliation{Zentrum f\"ur
Mathematische Physik und II. Institut f\"ur Theoretische Physik,
Universit\"at Hamburg, Luruper Chaussee 149, 22761 Hamburg,
Germany}

\author{Mattias N.\,R. Wohlfarth}
\email{mattias.wohlfarth@desy.de}
\affiliation{Zentrum f\"ur Mathematische Physik und II. Institut f\"ur Theoretische Physik, Universit\"at Hamburg, Luruper Chaussee 149,
22761 Hamburg, Germany}

%%%%%%%%%%%%%%%%%%%%%%%%%%%%%%%%%%%
\begin{abstract}
We reconsider both the global and local stability of solutions of continuously evolving dynamical systems from a geometric perspective. We clarify that an unambiguous definition of stability generally requires the choice of additional geometric structure that is not intrinsic to the dynamical system itself. While global Lyapunov stability is based on the choice of seminorms on the vector bundle of perturbations, we propose a definition of local stability based on the choice of a linear connection. We show how this definition reproduces known stability criteria for second order dynamical systems. In contrast to the general case, the special geometry of Lagrangian systems provides completely intrinsic notions of global and local stability. We demonstrate that these do not suffer from the limitations occurring in the analysis of the Maupertuis-Jacobi geodesics associated to natural Lagrangian systems.
\end{abstract}


\maketitle

%%%%%%%%%%%%%%%%%%%%%%%%%%%%%%%%%%%
\section{Introduction}
Continuously evolving dynamical systems universally appear as models in every branch of natural science. Their usefulness is determined by their predictivity which in turn hinges on the stability of their solutions. Given some uncertainty in the measured initial conditions, a physically meaningful system should offer control on the possible evolution of deviations from a given reference trajectory. Global control of late-time deviations is as important as a local understanding of stability. 
An extensive mathematical theory exists for the global stability of solutions of dynamical systems; the fundamental quantities in this theory are the Lyapunov exponents that measure exponential deviations~\cite{Oseledec:1968,Pesin:1977}. These exponents are often very hard to determine analytically, and various approaches for their calculation are used in the literature~\cite{Benettin:1976,Benettin:1980a,Benettin:1980b,Gozzi:1993tm,Caiani:1997yv,Casetti:2000gd,Terzic:2002uj}. Sometimes the stability of the same dynamical system is considered from different points of view, and the corresponding Lyapunov exponents are compared; this leads to some debate on the question of which point of view is preferable~\cite{Cerruti:1996,Bari:1997zz,Cipriani:1998zz,CuervoReyes,Cerruti:2008}. 
The local stability of solutions of dynamical systems on the other hand seems less well understood. To our knowledge, the existing mathematical theory is applicable only to second order differential systems. It was developed by Kosambi, Cartan and Chern~\cite{Kosambi,Cartan,Chern}, and covers as a special case the local stability analysis of geodesics on a Riemannian manifold via the Jacobi equation.

In this article we describe dynamical systems and their perturbations in a unified geometric language which is reviewed in section~\ref{sec_language}. This language has the advantage to specify very precisely the intrinsic geometric ingredients that define the dynamical system. It is worth emphasizing that the translation of dynamical systems into geometry is unambiguous. However, this is not true for the definition of stability. We will clarify in section~\ref{sec_stability} that stability in general is not an intrinsic notion but requires geometric structure not included in the definition of the dynamical system. This fact is not sufficiently appreciated in the literature. We will see that global Lyapunov stability entails the choice of seminorms on the vector bundle of perturbations. We will then propose a general definition of local stability applicable to any dynamical system. This notion of local stability requires the choice of a linear connection, and reduces to standard stability criteria for second order systems.
In contrast to the general case, Lagrangian systems offer intrinsic geometric structures to define global and local stability, as we will discuss in section~\ref{sec_Lagrange}. 
We will then consider natural Lagrangian systems in section~\ref{sec_natural} for which independent notions of stability exist based on the translation of solutions into geodesics via the Maupertuis-Jacobi principle. We point out that this detour suffers from a number of limitations which are nicely resolved in the direct intrinsic approach valid for all Lagrangian systems. We conclude in section~\ref{sec_conclusion}.

%%%%%%%%%%%%%%%%%%%%%%%%%%%%%%%%%%%
\section{Perturbations of dynamical systems}\label{sec_language}
In this section we review the basic geometric structure common to all dynamical systems and their perturbations. The geometric language used here will form the basis for our discussion of the stability of solutions in the following sections.

Fundamentally, deterministic dynamical systems can be introduced as formal rules that describe the evolution of points in some set $S$ with respect to an external, discrete or continuous time parameter running in another set $T$. More precisely, a dynamical system is a map
\begin{equation}\label{dynsyst}
\phi: T\times S  \rightarrow  S\,,\, (t,x)  \mapsto  \phi(t,x)
\end{equation}
which satisfies $\phi(t,\cdot)\circ\phi(s,\cdot)=\phi(t+s,\cdot)$ for all times $t,s\in T$. This bare definition must be enriched with additional structure if we want to model realistic dynamical systems. 

In the following we will therefore concentrate on the important class of continuously evolving dynamical systems for which $T=\mathbb{R}$. Geometrically, we focus on point sets $S$ that are differentiable manifolds, and assume that a vector field $X$ over $S$ is given:
\begin{equation}
X:S\rightarrow TS\,,\,p\mapsto X_p\in T_pS\,.
\end{equation}
The integral curves $\gamma_p$ of the vector field $X$ through a point $p\in S$ are defined by two conditions: that their tangent vector $\dot\gamma_p(t)=X(\gamma_p(t))$ agrees with the vector field at each point, and that $\gamma_p(0)=p$. For simplicity we assume that these integral curves are defined for all times. Then a dynamical system in the sense of (\ref{dynsyst}) is defined by the flow of the vector field $X$, i.e., by setting $\phi(t,p)=\gamma_p(t)$. 

\begin{figure}[ht]
\scalebox{0.7}{ \input{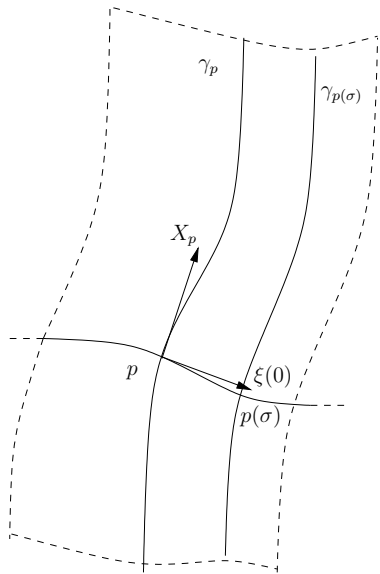} }
\caption{\label{figure1}\textit{Trajectories of dynamical systems and their perturbations.}}
\end{figure}

Any integral curve $\gamma_p$ of the flow vector field $X$ provides a solution trajectory of the dynamical system with initial condition $p$. Perturbations of $\gamma_p$ are studied by varying this initial condition. Any curve $p(\sigma)$ with $p(0)=p$ provides a continuous family of initial conditions which evolve under the dynamical system as $\phi(t,p(\sigma))=\gamma_{p(\sigma)}(t)$. All curves so obtained are continuous variations of $\gamma_p$; as a function of the two parameters $t$ and $\sigma$ we thus obtain a two-dimensional surface $\phi(t,p(\sigma))$ in $S$, see figure~\ref{figure1}. In consequence the two tangent vectors $X$ and $\xi(t,\sigma)=\partial_\sigma|_{\phi(t,p(\sigma))}$ commute, $[X,\xi]=0$. Along the reference solution $\gamma_p$ which lies in this surface at $\sigma=0$ we can now study the evolution of the tangent vector $\xi(t)=\partial_\sigma|_{\gamma_p(t)}$. Precisely this vector governs the linear stability of the solution~$\gamma_p$, i.e., its stability with respect to infinitesimal perturbations. Globally the perturbation vector field $\xi(t)$ along $\gamma_p$ is determined as the push-forward of an initial perturbation vector $\xi_0\in T_pS$ via
\begin{equation}\label{pushf}
\xi(t)=(\phi(t,\cdot))_*\xi_0\,.
\end{equation} 

To summarize, the dynamical systems $(S,X)$ that we will study in this paper have a very simple geometric structure: 
\begin{itemize}
\item they are defined in terms of a flow vector field $X$ over some differentiable manifold $S$; 
\item their solutions are integral curves of $X$; and
\item the linear perturbations of a solution $\gamma_p$ are vector fields $\xi$ along $\gamma_p$ for which $[X,\xi]=0$.
\end{itemize}
Choosing local coordinates $x^i$ on $S$, the dynamical system thus becomes a system of first-order differential equations over $S$ for the integral curves of the vector field $X$,
\begin{equation}
\dot x(t)^i=X^i(x(t))\,.
\end{equation} 
Expanding the commutator condition $[X,\xi]=0$ for the linear perturbation gives 
\begin{equation}
\dot{\xi}^i(t)=\left.\frac{\partial X^i}{\partial
x^j}\right|_{x(t)}\xi^j(t)\,.
\end{equation}

These local expressions seem to suggest that the above framework for dynamical systems would only allow for first order differential equations on a configuration space. However, this crucially depends on the choice of manifold $S$, its relation to the configuration space, and the type of flow vector field $X$. To see this we now discuss in the same geometric framework the case of second order dynamical systems which will play a major role in this paper (higher than second order construction require higher order tangent or jet bundles, see~\cite{Bucataru:2007b}). 

Hence consider a system whose configuration space is modeled by a smooth manifold $M$. Initial conditions for a second order system consist of an initial position $q\in M$ and an initial velocity $v\in T_qM$. Combining these into pairs $(q,v)$, one realizes that a natural geometric arena for second order systems is the tangent bundle $TM$. It is clear that not all vector fields over $TM$ can produce second order dynamics on $M$. Indeed, a second order system is defined by a vector field $X$ on $S=TM$ with the additional condition
\begin{equation}\label{secordcond}
  \pi_*X=u
\end{equation}
for all $(q,u)\in TM$, where $\pi_*$ is the push-forward of the projection map $\pi:TM\rightarrow M\,,\,(q,u)\mapsto q$. This condition defines $X$ as a so-called semispray; in local coordinates $(x^a,u^a)$ on $TM$ one finds $X=u^a \frac{\partial}{\partial x^a}+X_2^{a}\frac{\partial}{\partial u^a}$. This ensures that the integral curves of $X$, which satisfy
\begin{equation}
\dot{x}^a(t) = u^a(t)\,,\quad\dot{u}^a(t)= X_2^{a}(x(t),u(t)) \qquad\Rightarrow\qquad \ddot x^a(t)=X_2^a(x(t),\dot x(t))\,,
\end{equation}
represent a second order differential evolution on the configuration space $M$. In geometrical terms, the integral curves $\gamma$ of $X$ become natural lifts $\gamma(t)=(\tilde\gamma(t),\dot{\tilde\gamma}(t))$ to $TM$ of curves $\tilde\gamma$ in $M$.

Once second order systems on $M$ have been rewritten as (special cases of) first order systems on $S=TM$, perturbations of solutions can be analyzed as before. The perturbation
\begin{equation}\label{2perturbation}
\xi=\xi_1^a\frac{\partial}{\partial x^a}+\xi_2^{a}\frac{\partial}{\partial u^a}
\end{equation}
along a given solution $\gamma$ will now be a vector field over $TM$ with $[X,\xi]=0$. It is instructive to expand this condition in local coordinates $(x^a,u^a)$ on $TM$, which yields
\begin{equation}\label{2perturbation2}
\dot \xi_1^a = \xi_2^a\,,\quad \dot\xi_2^a=\frac{\partial X_2^a}{\partial x^b}\xi_1^b + \frac{\partial X_2^a}{\partial u^b}\xi_2^b \qquad\Rightarrow\qquad \ddot\xi_1^a = \frac{\partial X_2^a}{\partial x^b}\xi_1^b + \frac{\partial X_2^a}{\partial u^b}\dot \xi_1^b\,. 
\end{equation}
As expected, the perturbation equation can be read as a second order differential equation. Hence we see that the simple geometric framework for dynamical systems and  perturbations of solutions can be easily applied to higher order dynamics.

%%%%%%%%%%%%%%%%%%%%%%%%%%%%%%%%%%%
\section{Geometric notions of stability}\label{sec_stability}
In the previous section we have defined dynamical systems $(S,X)$ via vector fields $X$ and their associated flow on some differentiable manifold $S$. We have also presented the geometric language to define linear perturbations of specific solutions as connecting vector fields that commute with~$X$. In this section we discuss the issue of the stability of solutions under such linear perturbations, both globally and locally. We will find that stability generally is not intrinsic to a dynamical system, but requires additional geometric structure beyond $S$ and $X$ to be well-defined.

%%%%%%%%%%%%%%
\subsection{Global stability}
Consider a solution $\gamma_{p_0}$ of a dynamical system $(S,X)$. The linear perturbations of such a solution are described by tangent vector fields $\xi$ along $\gamma$ and satisfy $[X,\xi]=0$. Given an initial perturbation $\xi_0\in T_{p_0}S$, the dynamical system completely determines the evolution of $\xi$ along the solution $\gamma$ via the push-forward as in equation (\ref{pushf}). Since, intuitively, $\xi$ points to a neighboring solution, we wish to call $\gamma$ unstable, if any choice of $\xi_0$ results in exponential deviations of $\xi$ for late times. 

In order to implement this idea, we proceed in two steps. First we need to quantify the size of a perturbation $\xi$; so the tangent bundle $TS$ must be equipped with a family of (semi-)norms
\begin{equation}
\|\cdot\|_p: T_pS \rightarrow \mathbb{R}^+\,,\, \xi\mapsto\|\xi\|_p
\end{equation}
for each point $p$ of $S$. This is additional structure not provided in the definition of the dynamical system $(S,X)$. Second, once this geometric structure is available, we are in the position to define the so-called Lyapunov exponent \cite{Oseledec:1968,Pesin:1977}
\begin{equation}\label{Lyap}
\lambda (p_0,\xi_0)=\limsup_{t\rightarrow\infty}\frac{1}{t}\ln\|(\phi(t,\cdot))_*\xi_0\|_{\phi(t,p_0)}
=\limsup_{t\rightarrow\infty}\frac{1}{t}\ln\|\xi(t)\|_{\gamma_{p_0}(t)}
\end{equation}
for given initial point $p_0\in S$ and initial perturbation vector $\xi_0\in T_{p_0}S$. Exponential growth of a perturbation will result in positive Lyapunov exponent $\lambda$; therefore, a given trajectory identified by $p_0$ is defined as stable if and only if $\lambda(p_0,\xi_0)\le 0$ is non-positive for all initial perturbations~$\xi_0$. This is a definition of global stability because the Lyapunov exponents are evaluated for late times.

We note that the Lyapunov exponents and the notion of global  Lyapunov stability can be straightforwardly applied to any dynamical system formulated as a vector flow. For instance, in the case of a second order system over a configuration space $M$, which we have seen is a vector flow  over $S=TM$, the additional geometric structure needed to define global stability is a family of seminorms on the bundle $TS=TTM$. We will see this construction at work below.

Even though a family of seminorms is sufficient to define (\ref{Lyap}), many general results in the theory of Lyapunov exponents have been achieved by equipping $S$ with a Riemannian metric and considering the associated family of norms on $TS$. In this case, it can be shown that for a given point~$p$ the Lyapunov exponent $\lambda(p,\xi)$ can assume at most $n$ distinct values, ${\lambda_p^{(1)}<\dots<\lambda_p^{(s)}}$. Moreover, there exists a so called filtration $E_1\subset\dots\subset E_s=T_pS$ such that $\lambda(p,\xi)=\lambda_p^{(i)}$ for~${\xi\in E_i\setminus E_{i-1}}$. So initial perturbations from this subset of $T_pS$ result at late times in exponential deviations controlled by the corresponding Lyapunov exponent. Stability theory based on Lyapunov exponents is a vast subject; we will not detail its major achievements here but refer the interested reader e.g. to \cite{Benettin:1976,Benettin:1980a}. We only remark the following notable result: on a compact Riemannian manifold $S$, the supremum limit in (\ref{Lyap}) can be replaced by the ordinary limit  which, moreover, is finite and independent of the particular Riemannian metric defined on $S$.

One could suspect that, even though the Lyapunov exponent quantitatively depends on the specific choice of seminorms on $TS$, the qualitative notion of stability would not. 
The actual independence of Lyapunov exponents from a Riemannian metric on compact manifolds seems to favor this hypothesis. However, the following simple
example immediately reveals that this is not the case. Consider the inverted harmonic oscillator over $M=\mathbb{R}$ with Lagrangian
\begin{equation}\label{ex1}
L=\frac{1}{2}(u^2+\mu^2 x^2)
\end{equation}
for $\mu>0$. It is easy to see that the associated flow vector field on $TM$ is given by $X=(u,\mu^2 x)$. A generic perturbation of the solution $x=A e^{\mu t}+B e^{-\mu t}$  is given by ${\xi_1(t)=(a e^{\mu t}+be^{-\mu t})\partial_x}$, where $a$ and $b$ are arbitrary constants parametrizing the possible choices for the initial perturbation vector $\xi\in T_pTM$. The point $p$ at $t=0$ has coordinates $(x,u)=(A+B,\mu(A-B))$. The simple choice of seminorms induced on $TM$ by the Euclidean metric $g=dx\otimes dx$ gives the Lyapunov exponents
\begin{equation}
\lambda(A,B,a,b)=\lim_{t\rightarrow\infty}\frac{1}{t}\ln(a e^{\mu t}+be^{-\mu t})\,.
\end{equation}
The Lyapunov spectrum is then given by $\{-\mu,\mu\}$, where $-\mu$ is associated to the subspace $a=0$ of $T_pTM$ and $\mu$ to its complement. Therefore the solution is classified as unstable. However, why should we not consider the metric $g=(x^2+1)^{-1}dx\otimes dx$? Then the Lyapunov exponent
\begin{equation}
\lambda(A,B,a,b)=\lim_{t\rightarrow\infty}\frac{1}{t}\ln\frac{a e^{\mu t}+be^{-\mu t}}{(A e^{\mu t}+B e^{-\mu t})^2+1}
\end{equation}
vanishes (excluding the fixed point case $A=B=0$). The solution is now classified as stable in contradiction to the previous result.

We emphasize that the key point behind the Lyapunov notion of global stability of the trajectories of a dynamical system $(S,X)$ is the necessity for additional geometric  structure: seminorms on $TS$ that can measure the size of the perturbation. 

%%%%%%%%%%%%%%
\subsection{Local stability}
We now turn to a very intuitive proposal for the local stability of solutions of any dynamical system $(S,X)$. We wish to classify a solution $\gamma_{p_0}$  as locally stable if the acceleration between $\gamma_{p_0}$ and a neighboring solution is negative, i.e., if the curves are forced together. To implement this technically, we need to be able to calculate the second derivative of the perturbation vector field $\xi$ that points to the neighboring solutions. The only derivative that is available without introducing further structure is the Lie derivative, but already the first derivative $\mathcal{L}_X\xi=[X,\xi]=0$ vanishes. So we need to require another derivative. 

The definition of local stability again proceeds in two steps. First, we introduce a linear connection $\nabla_XY$ that sends a pair of vector fields $X,Y$ over $S$ to another vector field over $S$. Using the connection, one finds that the second covariant derivative of $\xi$ along the flow $X$ is a linear operator on $TS$ which can be written as
\begin{equation}\label{Rlin}
\nabla_X\nabla_X\xi = \left[\nabla_X\left(T(X,\cdot)+\nabla_\cdot X\right)+\left(T(X,\cdot)+\nabla_\cdot X\right)^2\right]\cdot\xi = \tilde R(X)\cdot \xi\,,
\end{equation} 
where $T(X,Y)=\nabla_XY-\nabla_YX-[X,Y]$ is the torsion tensor of $\nabla$, and both $T(X,\cdot)$ and $\nabla_\cdot X$ are linear operators on $TS$. Second, we define a solution $\gamma_{p_0}$ of the dynamical system as locally stable if all eigenvalues of the linear operator $\tilde R$ evaluated along $\gamma_{p_0}$ are non-positive.

To our knowledge this definition of local stability, which can be applied to dynamical systems of any order, has not appeared in the literature. However, we will now show that our proposal has a very nice relation to KCC stability theory as developed by Kosambi, Cartan and Chern \cite{Kosambi,Cartan,Chern} for second order dynamical systems over a configuration space manifold~$M$. We have already seen that these systems are defined as vector flows over the tangent bundle~$S=TM$ in terms of a semispray~$X$. Without loss of generality we may use local coordinates $x^A=(x^a,u^a)$ on $S$ to write 
\begin{equation}\label{sem}
X = u^a\frac{\partial}{\partial x^a} -2G^a(x,u)\frac{\partial}{\partial u^a}\,.
\end{equation}

To discuss the geometry of the dynamical system so defined, KCC introduce a non-linear connection $N$ on $M$ with coefficients $N^a{}_b = \frac{\partial G^a}{\partial u^b}$. This can be understood in terms of a dynamical covariant derivative $\nabla^N$: for two vector fields $v,w$ over $M$,
\begin{equation}
\nabla^N_vw = \left[v^a\partial_a(w^b) + N^b{}_a(x,v) w^a\right]\frac{\partial}{\partial x^b}\,.
\end{equation}
(In the special case of a linear connection one simply has $N^a{}_b(x,v)=\Gamma^a{}_{bc}(x)v^c$.) Using $\nabla^N$ one can rewrite the second order equation in~(\ref{2perturbation2}), which governs the evolution of a perturbation vector~$\xi_1$ over $M$, in the form
\begin{equation}\label{KCCs}
\nabla^N_{\dot x}\nabla^N_{\dot x}\xi_1 = P(\dot x)\cdot \xi_1
\end{equation} 
for a linear operator $P$ on $TM$ the components of which may be found e.g. in \cite{Harko:2008ak}:
\begin{equation}\label{Pcomp}
P^a{}_b = -2\frac{\partial G^a}{\partial x^b} -2 G^c \frac{\partial N^a{}_b}{\partial u^c} + u^c \frac{\partial N^a{}_b}{\partial x^c} +N^a{}_cN^c{}_b\,.
\end{equation}
A solution $\gamma_{p_0}$ of the dynamical system is called KCC-stable if all eigenvalues of~$P$ evaluated along $\gamma_{p_0}$ are non-positive.

Now to define local stability of the system $(S=TM,X)$ with $X$ provided by (\ref{sem}) according to our proposal, we need to specify a linear connection $\nabla$ over $S=TM$ instead of a non-linear connection over $M$. For this purpose we note that for every non-linear connection $N$ on $M$ exists an associated compatible linear connection $\nabla$ on $TM$, which is said to be of Berwald type. Technical details and the global form of this linear connection are given in~\cite{Bucataru:1999}. The components of $\nabla$ are defined by $\nabla_AE_B=\Gamma^C_{BA}E_C$ with respect to a basis $(E_A)$ of $TTM$. Choosing the basis $(E_A)=(\frac{\partial}{\partial x^a}-N^b{}_a\frac{\partial}{\partial u^b},\frac{\partial}{\partial u^a})$ adapted to~$N$, the only non-vanishing connection components are
\begin{equation}\label{Bconn}
\Gamma^c_{ba}=\frac{\partial^2G^c}{\partial u^b\partial u^a}\,,\quad \Gamma^{\bar c}_{\bar b a}= \frac{\partial^2G^c}{\partial u^b\partial u^a}\,.
\end{equation}
In the same basis, we now calculate the linear operator $\tilde R(X)$ on $TS=TTM$ defined in~(\ref{Rlin}), the eigenvalues of which are relevant for our proposal for local stability. The result has the form
\begin{equation}
\tilde R^A{}_B = \left[\begin{array}{cc} R_1{}^a{}_b+P^a{}_b & -2 (\bar\partial \epsilon)^a{}_{\bar b} \\ R_2{}^{\bar a}{}_b & R_1{}^{\bar a}{}_{\bar b}+P^{\bar a}{}_{\bar b}  \end{array} \right]
\end{equation}
in terms of the deviation tensor $P$ that appears in the stability equation (\ref{KCCs}) of KCC theory, of $\epsilon^a = 2G^a - N^a{}_bu^b$, of $(\bar \partial\epsilon)^a{}_b=\frac{\partial\epsilon^a}{\partial u^b}$, and of two more tensors $R_1$ and $R_2$. We only need to know that $\bar\partial\epsilon=0$ implies $R_1=0$; in this case the eigenvalues of $\tilde R$ are precisely those of $P$, so our definition of local stability reduces to the KCC definition. Note that $\bar\partial\epsilon=0$ is solved by 
\begin{equation}\label{Gsol}
G^a(x,u)=G^a_{(2)}(x,u)+\frac{1}{2}\epsilon^a(x)
\end{equation}
where $G^a_{(2)}(x,u)$ is homogeneous in $u$ of degree two. We will see that semisprays with such $G^a$ arise from natural Lagrangian systems. So our criterion for local stability agrees with the KCC criterion for all natural Lagrangian systems.

The fact that local stability of a second order system can be equivalently understood in terms of a nonlinear connection on $M$ or in terms of a linear connection on $TM$ as we propose was already shown for~$\epsilon=0$ in~\cite{Bucataru:1999}, but in a different way. This is in fact the only case where the KCC stability treatment seems to be natural, because the second order system then takes the autoparallel form $\nabla^N_{\dot x}\dot x=0$. But also the Berwald type connection satisfies $\nabla_XX=0$. For all other cases we do not consider a non-linear connection a particularly natural structure. 
Another point in favor of our linear connection proposal is the fact that it is naturally defined for any dynamical system $(S,X)$, not only for those related to second order dynamics.

We emphasize that the key point behind the notion of local stability of the trajectories of a dynamical system $(S,X)$ here proposed is the necessity for additional geometric  structure: a connection $\nabla$ on $TS$ that allows to measure the acceleration between nearby curves. 

%%%%%%%%%%%%%%
\subsection{Geodesics}
Both the definitions of global and local stability of the solutions of generic dynamical systems~$(S,X)$ require the choice of additional geometric structure: in the first case seminorms have to be chosen on~$TS$ to measure the size of the perturbation; in the second case a connection on~$TS$ has to be chosen to measure local accelerations. 

However, some dynamical systems may provide natural choices. In the case of geodesics on a Riemannian manifold $(M,g)$, which can be seen as a second order dynamical system $(S=TM,X)$, natural choices exist for the seminorms as well as for a connection on $TS=TTM$. 

The seminorms for the definition of global Lyapunov stability can be constructed from any one of the well-known lifts of the metric $g$ to the tangent bundle which are discussed in detail in~\cite{Yanobook}. As an example we may choose seminorms
\begin{equation}
\|\xi\|_p^2=(g^V)_p(\xi,\xi)
\end{equation}
on $TTM$ by employing the vertical lift of $g$ to $TM$ which is defined by $g^V(X,Y)=g(\pi_*X,\pi_*Y)$ for vector fields $X,Y$ on $TM$. This construction relates instability to the exponential divergence of trajectories purely in the configuration space. In the notation of (\ref{2perturbation}) the degenerate bilinear form $g^V$ projects out the components $\xi_2^a$. Indeed, for a geodesic $\tilde\gamma:\mathbb{R}\rightarrow M$ with initial condition $(q,v)=p \in TM$ and initial perturbation $(\xi_1,\xi_2)=\xi\in T_pTM$, the Lyapunov exponents are simply calculated as
\begin{equation}
\lambda(p,\xi) = \lim_{t\rightarrow\infty} \frac{1}{2t}\ln g_{\tilde\gamma(t)}(\xi_1(t),\xi_1(t))\,.
\end{equation}

Affinely parametrized geodesics with tangent vector field $\dot x$ over $M$ obey the equation $\nabla^{LC}_{\dot x}\dot x=0$ where $\nabla^{LC}$ is the Levi-Civita linear connection over $M$ determined by the metric $g$. As discussed in the previous section, there is a linear connection $\nabla$ over $TM$ naturally associated to $\nabla^{LC}$ (which now plays the role of $N$). The connection $\nabla$ of Berwald type is nothing else but the horizontal lift $\nabla=\nabla^H$ of~$\nabla^{LC}$ to~$TM$; see~\cite{Yanobook} for details. According to our definition, local stability of a geodesic requires that the eigenvalues of $\tilde R(X)$ as defined in~(\ref{Rlin}) should be non-positive. From the calculation made above in the comparison to KCC theory we hence deduce that local stability becomes equivalent to the eigenvalues of $P(\dot x)^a{}_b = - R^a{}_{pbq}\dot x^p\dot x^q$ being non-positive. Considering the geodesic deviation equation
\begin{equation}\label{Jacobi}
(\nabla^{LC}_{\dot x}\nabla^{LC}_{\dot x}\xi_1)^a = - R^a{}_{pbq}\dot x^p\dot x^q \xi_1^b
\end{equation}
for a simple perturbation $[\dot x,\xi_1]=0$ on $M$ then tells us that our notion for the local stability of geodesics naturally reduces to the standard definition of geodesic stability.

So geodesics admit natural definitions of global as well as local stability. In the following section we will show that also Lagrangian systems offer intrinsic notions of global and local stability. 
 
%%%%%%%%%%%%%%%%%%%%%%%%%%%%%%%%%%%
\section{Stability analysis of Lagrangian systems}\label{sec_Lagrange}
In the previous sections we have introduced all necessary ingredients for a discussion of the stability of solutions of Lagrangian systems on some configuration space manifold $M$. In order to apply the geometric machinery we will first formulate these systems as vector flows over~$TM$. Then we will demonstrate the existence of natural notions of stability that do not require further information apart from that given by the Lagrangian system itself. 

%%%%%%%%%%%%%%
\subsection{Geometric formulation}\label{Lageo}
A Lagrangian system is a pair $(M,L)$ of a configuration space manifold $M$ and a Lagrangian function $L:TM\rightarrow \mathbb{R}$. Locally this is a function $L(x,u)$ of  coordinates $x^a$ on~$M$ and coordinates~$u^a$ on $T_xM$ that give the components of a vector $u$ in the coordinate-induced basis $\frac{\partial}{\partial x^a}$. The trajectories of the Lagrangian system are the curves $\gamma:\mathbb{R}\rightarrow M$ with tangent vectors $\dot \gamma$ for which the variation of the action integral vanishes,
\begin{equation}
 0=\delta S[\gamma]=\delta\int_{t_1}^{t_2}
 L(\gamma(t),\dot{\gamma}(t))\,dt\,.
\end{equation}
In the variation the curve's endpoints $\gamma(t_1)=p_1$ and $\gamma(t_2)=p_2$ are held fixed. Once coordinates have been chosen, solutions $x^a(t)$ must satisfy the Euler-Lagrange equations
\begin{equation}
\frac{d}{dt}\frac{\partial L}{\partial \dot{x}^a}-\frac{\partial L}{\partial x^a}=0\,.
\end{equation}

As any second order dynamical system, a Lagrangian system can be described as a vector flow $X$ on $S=TM$. Since this formulation is not so well-known, we present the relevant steps following~\cite{Crampin}. First, one defines the Liouville vector field $\Delta=u^a\frac{\partial}{\partial u^a}$ on $TM$. This produces the energy function $E:TM\rightarrow M$ by $E=\Delta(L)-L$. Now recall that a vector field $Y$ on $TM$ is said to be vertical if it is tangent to the fiber, $\pi_* Y=0$. Moreover, a vector $w=w^a\frac{\partial}{\partial x^a}$ on $M$, has an associated vertical lift $w^V= w^a\frac{\partial}{\partial u^a}$ to $TM$. Combining the vertical lift and $\pi_*$ defines a $(1,1)$ tensor field $S$ on $TM$ by $S(Y)=(\pi_* Y)^V$; in coordinates, $S=\frac{\partial}{\partial u^a}\otimes dx^a$. With these ingredients, we can construct the two-form $\omega=d(dL\circ S)$ on $TM$. As a final step, the vector field $X$ generating the Lagrangian flow is the unique vector field $X$ solving
\begin{equation}\label{ELgeo}
i_X\omega=-dE\,. 
\end{equation}
This construction is easily understood in a local chart $(x,u)$ of $TM$. Simple algebra shows that
\begin{eqnarray}
 dE&=&\Big(\frac{\partial^2 L}{\partial x^a \partial u^b}u^b-\frac{\partial L}{\partial x^a}\Big)dx^a+\frac{\partial^2 L}{\partial u^a\partial u^b}u^b du^a\,,\nonumber\\
 \omega&=&\frac{\partial^2 L}{\partial x^a \partial u^b} dx^a\wedge dx^b+\frac{\partial^2 L}{\partial u^a \partial u^b}du^a\wedge dx^b\,.
\end{eqnarray}
Writing $X=X_1^a\frac{\partial}{\partial x^a}+X_2^a\frac{\partial}{\partial u^a}$, condition (\ref{ELgeo}) is equivalent to
\begin{equation}\label{Lvec}
X_1^a = u^a\,,\quad
\frac{\partial^2 L}{\partial u^a \partial u^b} X_2^a+\frac{\partial^2 L}{\partial x^a \partial u^b}X_1^a-\frac{\partial L}{\partial x^b}=0\,.
\end{equation}
This shows that the vector field $X$ on $TM$ is a semispray, as needed for a second order dynamical system. The integral curves of $X$ indeed solve the Euler-Lagrange equations.

%%%%%%%%%%%%%%
\subsection{Intrinsic notions of stability}
Within the geometric picture of Lagrangian systems $(M,L)$ as dynamical systems $(S=TM,X)$ we are now in the position to ask for the stability of solutions. We will discuss global and local stability in turn.

The definition of global stability via Lyapunov exponents requires the specification of a family of seminorms on the tangent bundle $TS$. A Lagrangian system provides us with an intrinsic object that can be used for this purpose, see e.g. \cite{Bucataru:2007a}: the generalized Lagrange metric
\begin{equation}\label{Lmet}
g = \frac{\partial^2L}{\partial u^a\partial u^b} dx^a\otimes dx^b\,.
\end{equation} 
Although the coefficients of $g$ depend on the fiber coordinates $u$ for generic Lagrangians~$L$, the generalized Lagrange metric transforms under coordinate transformations of $S=TM$ as a $(0,2)$-tensor on $M$. Regularity of the Lagrangian corresponds to $\textrm{rank }g=\dim M$. The situation for the Lagrangian system is now very similar to that of geodesics on a Riemannian manifold. Instead of lifting the Riemannian metric to $TM$, we simply need to extend the generalized Lagrange metric~$g$ to $TM$. As for lifts, there are several possibilities. As an example we consider the case, where $g$ as displayed above is simply read as a tensor over $TM$. For a reference solution $\tilde\gamma:\mathbb{R}\rightarrow M$ with initial condition $(q,v)=p \in TM$ and initial perturbation $(\xi_1,\xi_2)=\xi\in T_pTM$, the Lyapunov exponents are then simply calculated as
\begin{equation}\label{LL}
\lambda(p,\xi) = \lim_{t\rightarrow\infty} \frac{1}{2t}\ln g_{\tilde\gamma(t)}(\xi_1(t),\xi_1(t))\,.
\end{equation}
This is of the same form as the formula for geodesics where we used the vertical lift of the Riemannian metric. So this choice of seminorms measures the exponential deviation of perturbations purely within the configuration space $M$.

To define the local stability of solutions of $(M,L)$ we need to specify a linear  connection acting on vector fields on $S=TM$. Again the Lagrangian system provides us with an intrinsic choice. We may simply use the Berwald type linear connection $\nabla$ on $TM$ which is associated to any given semispray. The relevant quantities $G^a$ and $N^a{}_b = \frac{\partial G^a}{\partial u^b}$ needed to determine this connection follow by combining (\ref{sem}), (\ref{Lvec}) and the definition of the generalized Lagrange metric:
\begin{equation}\label{GLag}
G^a = \frac{1}{2} g^{ab}\frac{\partial^2L}{\partial u^b\partial x^c}u^c-\frac{1}{2}g^{ab}\frac{\partial L}{\partial x^b}\,. 
\end{equation}
In the basis $(E_A)=(\frac{\partial}{\partial x^a}-N^b{}_a\frac{\partial}{\partial u^b},\frac{\partial}{\partial u^a})$ of $TS$ the connection $\nabla$ is then provided by the connection coefficients~(\ref{Bconn}).

Both the global and local stability of solutions of a Lagrangian system $(M,L)$ can be defined intrinsically, without the introduction of further geometric structure. This makes essential use of the fact that the Lagrangian system can be rewritten in the dynamical system language as a vector flow $X$ over $S=TM$.  

%%%%%%%%%%%%%%%%%%%%%%%%%%%%%%%%%%%
\section{Natural Lagrangian systems}\label{sec_natural}
The solutions of natural Lagrangian systems can be translated using the Maupertuis-Jacobi principle to geodesics on an associated Riemannian manifold. We have seen that both global and local stability can be easily defined for geodesics, and this fact is often used in the literature to analyze the stability of natural Lagrangian systems. The main purpose of this section is to clarify the differences between the intrinsic notions of stability discussed in the previous section and the notions of stability obtained from the translation principle. It will turn out that the intrinsic notions of stability have considerable advantages.

%%%%%%%%%%%%%%
\subsection{Intrinsic stability}
Before we translate natural Lagrangian systems into geodesics, we specialize the intrinsic notions of stability developed for general Lagrangian systems to this case. A natural Lagrangian system is defined by a Lagrangian function $L:TM\rightarrow M$ which is quadratic in the velocities:
\begin{equation}\label{natL}
L(x,u) = \frac{1}{2}k_{ab}(x)u^au^b - V(x)\,.
\end{equation}
This corresponds to a dynamical system $(S=TM,X)$ for which the flow vector field $X$ over $TM$ is determined according to the general expression~(\ref{Lvec}), and in terms of the Levi-Civita connection of the kinetic energy metric $k$, as
\begin{equation}
X_1^a=u^a\,,\quad X_2^a= -2G^a = -\Gamma_{(k)}{}^a{}_{bc}u^bu^c - \nabla_{(k)}{}^aV\,.
\end{equation}

The intrinsic notion of global stability is based on the generalized Lagrange metric $g$ of~(\ref{Lmet}) which here agrees with the kinetic energy metric, $g=k$. Assuming a regular Lagrangian and positive energy amounts to the requirement of positive definiteness of $k$. The second order evolution equation of perturbations $\xi_1$ can be rewritten as
\begin{equation}\label{soform}
\Big(\nabla^{(k)}{}_{\dot x}\nabla^{(k)}{}_{\dot x}\,\xi_1\Big)^a =
- \Big(R_{(k)}{}^a{}_{pbq}\dot x^p\dot x^q + \nabla_{(k)}{}^a\nabla^{(k)}{}_{b}V\Big)\xi_1^b\,.
\end{equation}
The calculation of Lyapunov exponents with respect to some extension of $k$ to $TM$ is unproblematic and follows~(\ref{LL}). 

The intrinsic notion of local stability is linked to the Berwald type connection on $S=TM$. Note that the natural Lagrangian system has a flow vector field $X_2^a=-2G^a$ of the type~(\ref{Gsol}), so our general calculation above shows that local stability becomes equivalent to the KCC definition. In practice, therefore, local stability means non-positive eigenvalues of the endomorphism $P$ defined in~(\ref{Pcomp}). Specializing to the case here we find
\begin{equation}\label{Pnat}
P^a{}_b =  -R_{(k)}{}^a{}_{pbq}u^pu^q - \nabla_{(k)}{}^a\nabla^{(k)}{}_b V\,, 
\end{equation}
where $u$ has to be evaluated for the tangent vectors $\dot x$ along a reference solution. This is the same operator as the one we found governing the evolution of perturbations.

Reconsider the example of the inverted harmonic oscillator of equation~(\ref{ex1}). The generalized Lagrange metric now gives a preferred intrinsic choice of seminorms in terms of $g=dx\otimes dx$, thus leading to the expected global Lyapunov instability of the system. For local stability we use the fact that our stability criterion and the KCC criterion agree, since we consider a natural Lagrangian system. From~(\ref{Pnat}) we find $P=\mu^2>0$. This shows the local instability of all solutions under this intrinsic choice of connection.

%%%%%%%%%%%%%%
\subsection{Riemannian translation}
Geodesics on a Riemannian manifold with fixed affine parametrization can be viewed as a Lagrangian system. The converse is also true for natural Lagrangian systems. We now state a version of the Jacobi-Maupertuis principle, including indefinite kinetic energy metrics which occur e.g. in scalar field cosmology~\cite{Townsend:2004zp,Elias:2007uy}; then we will use this translation principle to investigate how the geodesic stability and the intrinsic stability of natural Lagrangian systems compare.

Let~$(M,L)$ be a natural Lagrangian system (\ref{natL}) with non-degenerate quadratic form $k$. Then the solution trajectories of fixed energy
\begin{equation}\label{Hamilton}
E = \frac{1}{2}k_{ab}(x)\dot x^a\dot x^b + V(x)
\end{equation}
of the Lagrangian system coincide with the geodesics of the metric manifold $(M,g)$ with
\begin{equation}\label{MJmetric}
g_{ab}=C\,|E-V(x)|\,k_{ab}
\end{equation}
for constant $C>0$ (the conventional choice is $C=2$). The time parameter $t$ of the Lagrangian system is related to the affine parameter $\tau$ of the geodesics, which we define by $|g_{ab}\frac{dx^a}{d\tau} \frac{dx^b}{d\tau}|=1$, as
\begin{equation}
\left|\frac{dt}{d\tau}\right|=\frac{1}{\sqrt{2 C}\,|E-V(x)|}\,.
\end{equation}

To see why the Jacobi-Maupertuis principle holds, we write the Euler-Lagrange equations of motion of the Lagrangian system as
\begin{equation}\label{eqLagr}
\ddot{x}^a+\Gamma_{(k)}{}^a{}_{bc}\,\dot{x}^a\dot{x}^b+k^{ab}\partial_b
V=0\,.
\end{equation}
Now consider geodesics on a metric manifold $(M,g)$ with the
conformally rescaled metric ${g_{ab}=\sigma^2 k_{ab}}$, parametrized
in terms of an arbitrary non-affine parameter also called~$t$.
These geodesics satisfy the equation
\begin{equation}
\ddot x^a + \Gamma_{(g)}{}^a{}_{bc}\dot x^a \dot x^b -\frac{1}{2}\dot x^a \frac{d}{dt}\ln |g_{pq}\dot x^p\dot x^q| = 0\,,
\end{equation}
now using the Christoffel symbols $\Gamma_{(g)}$ of the
rescaled metric $g$, and may be rewritten in terms of the original
kinetic energy metric $k$ and the conformal factor $\sigma$ as
\begin{equation}
\ddot x^{a}+\Gamma_{(k)}{}^a{}_{bc}\dot x^a \dot x^b - (k_{pq}\dot x^p
\dot x^q)\,k^{ab}\partial_b\ln|\sigma| + \dot x^a\frac{d}{dt}\ln\frac{|\sigma|}{|k_{pq}\dot x^p
\dot x^q|^{1/2}} = 0\,.
\end{equation}
This equation is equivalent to the Euler-Lagrange equations~(\ref{eqLagr}) provided that
\begin{equation}\label{condi}
k^{ab}\partial_b V + (k_{pq}\dot x^p \dot x^q)\,k^{ab}\partial_b\ln|\sigma|
- \dot x^a\frac{d}{dt}\ln\frac{|\sigma|}{|k_{pq}\dot x^p \dot x^q|^{1/2}} = 0\,.
\end{equation}
The contraction of this condition with $\dot x^c k_{ca}$ is easily integrated with integration constant~$E$, yielding the Hamiltonian constraint~(\ref{Hamilton}). We solve this for $k_{pq}\dot x^p\dot x^q$ and substitute back into the above condition for the equivalence of the geodesic and Euler-Lagrange equations. Rearranging the terms we obtain
\begin{equation}
k^{ab}\partial_b \ln \frac{\sigma^2}{|E-V|} - \frac{\dot x^a}{2(E-V)} \frac{d}{dt}\ln \frac{\sigma^2}{|E-V|}= 0\,.
\end{equation}
This is satisfied for all possible trajectories only if $\sigma^2 = C\,|E-V|$ for positive constant~$C$, which demonstrates the first part of the claim. Finally, $|g_{ab}\frac{dx^a}{d\tau}\frac{dx^b}{d\tau}|=1$ defines an affine parameter $\tau$ along the geodesics of $(M,g)$. We may now use the Hamiltonian constraint and our result for $\sigma^2$ to show that $(dt/d\tau)^2=\sigma^{-2}/2$ relates the affine parameter to the Lagrangian time.

%%%%%%%%%%%%%%
\subsection{Geodesic stability}
Using the translation principle, the stability of solutions of natural Lagrangian systems can now be discussed in terms of the stability of the corresponding geodesics. It turns out that this approach has a number of limitations that are not always appreciated in the literature:
\begin{itemize}
\item the boundaries of the Riemannian manifold;
\item the absence of perturbations that modify energy and the change to affine time; 
\item the problem of one-dimensional systems.
\end{itemize}
These points equally impact the discussion of global and local stability.

\vspace{6pt}\paragraph*{Boundaries of the Riemannian manifold.}
The Maupertuis-Jacobi principle translates the solutions of a natural Lagrangian system $(M,L)$ with given energy $E$ into geodesics of an associated $n$-dimensional metric manifold $(M,g^E)$ with metric $g_E=\sigma^2 k$, see (\ref{MJmetric}). This Riemannian metric has a singular boundary at all points where the conformal factor $\sigma^2$ vanishes, i.e., for $V(x)=E$, even assuming regularity of the kinetic energy metric $k$. The Ricci scalar
\begin{equation}
R_{(g_E)}=\sigma^{-2}\big(R_{(k)}-2(n-1)\,\square_{(k)}\ln\sigma-(n-2)(n-1)\nabla_{(k)}{}^a\ln\sigma\,\nabla^{(k)}{}_a\ln\sigma\big)\,,
\end{equation}
generically diverges for a vanishing $\sigma$. This means that the boundary singularity is a curvature singularity, not an artifact due to a possibly poor choice of coordinates. 

The fact that the Riemannian manifold used to model the Lagrangian system has a singular boundary can give rise to difficulties in the discussion of the equivalent dynamical system, see e.g.~\cite{Hrycyna:2005qp}. All solutions for which the velocity vanishes at some time parameter will hit this boundary; therefore their evolution cannot be determined in the geodesic picture, even though it is perfectly well-defined on the pure dynamical side. Consequently, neither the global nor the local stability of solutions hitting the boundary can be determined: the geodesics cannot be extended outside a limited range of their proper parameter, so the late time limit that needs to be taken for the Lyapunov exponents is ill-defined; also local stability cannot be checked along a complete trajectory. Moreover note that fixed point solutions, with vanishing velocity everywhere, are not tractable in the geodesic picture at all. This strikingly contrasts the dynamical system analysis, in which stability of fixed points is well understood.

\begin{figure}[ht]
\scalebox{0.7}{ \input{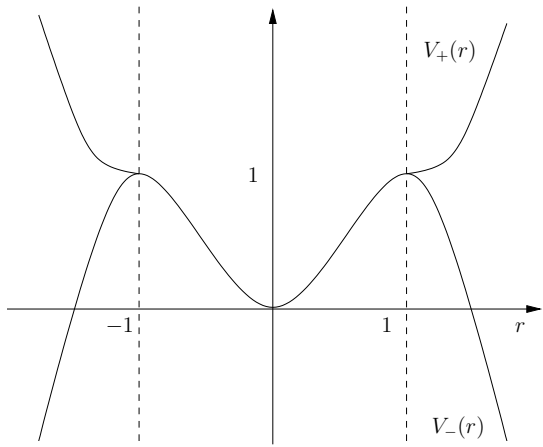} }
\caption{\label{figure2}\textit{Potentials of different dynamical systems mapped to the same Riemannian manifold.}}
\end{figure}

The relevance of these issues is easily illustrated already in simple radially symmetric systems
\begin{equation}
L=\frac{1}{2}(\dot{x}^2+\dot{y}^2)-V(r)
\end{equation}
with $r=(x^2+y^2)^{1/2}$. First, consider the harmonic oscillator potential $V(r)=r^2$ and solutions of energy $E=1$. In this case, the Riemannian manifold will be restricted to $\{(x,y)\,|\,0\leq r<1\}$ with metric $g=2(1-r^2)(dx\otimes dx+dy\otimes dy)$. The Ricci scalar is given by $R=2(1-r^2)^{-3}$, so that the boundary is indeed singular. One may check that all radial geodesics, characterized in polar coordinates by
$\dot{\phi}=0$, reach the boundary in finite proper time. Therefore the manifold is also geodesically incomplete. These conclusions hold in full generality for similar potentials, too. To illustrate the consequences for the stability analysis we consider the potentials
\begin{equation}
V_{\pm}(r)=2r^2-r^4\pm 2\Theta(r^2-1)(r^2-1)^2\,,
\end{equation}
where $\Theta$ is the Heaviside step function, see figure \ref{figure2}. Solutions with $E=1$ and initial position close to $r=0$ are mapped to the same Riemannian manifold for both potentials. So the differences of the two dynamical systems cannot be resolved in their geometrical image. In particular, the stability of radial solutions under small perturbations is (under any reasonable intuitive idea of stability) dramatically different in the two systems, which the geodesic picture is unable to reveal.

On the positive side, we remark that in complex systems solutions reaching the boundary of the manifold are the exception more than the rule, at least if the kinetic energy metric is positive definite. The set of initial conditions for solutions reaching the boundary will be very restricted. Solutions with generic initial conditions are expected to be confined to the interior of the manifold. In the examples above, only radial geodesics reach the boundary since an arbitrarily small angular momentum already prevents points of vanishing velocity along the trajectory. 

\vspace{6pt}\paragraph*{Perturbations that modify energy and the change to affine time.}
The energy of the solutions of a Lagrangian system directly enters the
definition of the metric of the corresponding Riemannian manifold. Hence a generic Lagrangian system cannot be mapped completely into a single metric manifold. Note that two metric manifolds $(\mathcal{M},g_{E_1})$ and $(\mathcal{M},g_{E_2})$
corresponding to different choices of fixed energy do not possess isometric regions, which makes it impossible to use part of a given manifold to discuss dynamics at different energy. For the discussion of both global or local stability this means that only perturbations leaving the energy fixed can be considered. From the Lagrangian point of view an initial perturbation is represented by a vector $\xi$ in the $2n$-dimensional space $T_pTM$, where $p\in TM$ provides initial conditions for the reference solution. A perturbation modifies the energy of the trajectory unless $\xi(E)=dE(\xi)=0$, where $E$ is the energy function defined in section~\ref{Lageo}. We can assume $dE\neq 0$, because the system otherwise would be in a fixed point, so  $dE(\xi)=0$ restricts $\xi$ to a $(2n-1)$-dimensional subspace of $T_pTM$. In the geodesic translation we therefore expect to have control only over this subspace. 

Another point to note is that the preferred role of the Lagrangian time is lost in translation from the Lagrangian picture to the geodesic picture. In the latter the only geometrically natural parameter is the affine parameter along the geodesics. That this has impact on the discussion of stability becomes clear in simple examples, where time rescalings may render unstable solutions stable and vice versa.

The non-existence of energy-changing perturbations and the loss of Lagrangian time reduce the control of stability of natural Lagrangian systems when discussed in the geodesic picture. This fact is nicely reflected in the reduction of the number of relevant exponents in the Lyapunov spectrum. To see this note that the evolution equation for the perturbation vector $\xi_1$ in (\ref{2perturbation2}) can be recast as the Jacobi equation (\ref{Jacobi}) using the Levi-Civita connection of $g_E$. In a parallely transported orthonormal basis $(e_a)=(e_0=\dot x,e_\alpha)$ with $\nabla^{LC}_{\dot x}e_a=0$ the Jacobi equation becomes 
\begin{equation}\label{geomsplit}
\ddot{\xi}_1^0=0\,,\quad 
\ddot{\xi}_1^\alpha =-R^\alpha{}_{0\beta 0}\xi_1^\beta\,.
\end{equation}
Choosing as initial perturbation $\dot\xi_1^0=0$, $\xi_1^\alpha=0$, $\dot{\xi}_1^\alpha=0$ gives the solution $\xi_1^0=\textrm{const}$, $\xi_1^\alpha=0$, and the corresponding Lyapunov exponent clearly vanishes. The perturbation corresponding to this exponent is an infinitesimal shift along the geodesic. The choice of initial conditions $\xi_1^0=0$, $\xi_1^\alpha=0$, $\dot \xi_1^\alpha=0$ gives the solution $\xi_1^0=\textrm{const}\cdot\tau$, $\xi_1^\alpha=0$, and hence a second vanishing Lyapunov exponent. In this case, the initial perturbation corresponds to an infinitesimal shift of the initial velocity along the geodesic. As a consequence, only $2n-2$ Lyapunov exponents in the spectrum remain significant.

\vspace{6pt}\paragraph*{One-dimensional systems.}
While it is often argued that the geodesic translation is useful for complex systems~\cite{Casetti:2000gd}, one must emphasize that it completely breaks down for simple one-dimensional systems. We already noted above that one obtains two vanishing Lyapunov exponents for any translated system; for $n=1$ this means that all Lyapunov exponents vanish. So global stability cannot distinguish arbitrary one-dimensional systems. The situation is similarly bad for local stability. All one-dimensional Riemannian manifolds are flat, so the eigenvalues of the Riemann tensor are zero. But then also local stability cannot distinguish arbitrary dynamical systems. Both concepts become meaningless for one-dimensional systems.

\vspace{6pt}After analyzing the different aspects of the geometric picture, we conclude that the Riemannian way to define stability suffers from a number of limitations. General 
dynamical systems cannot be mapped into a unique Riemannian manifold, which leads to the presence of singular boundaries that may spoil the method of any predictive power. Perturbations that modify the energy cannot be discussed at all. Systems with a small number of degrees of freedom are affected more severely; the method is not applicable to the one-dimensional case. 

None of these limitations apply to the intrinsic definitions of stability. The generalized Lagrange metric is just $k$ without any conformal factor, so that no singular boundaries can occur. Full control on perturbations that change energy is expected; indeed, the additional contribution from the potential in~(\ref{soform}) in contrast to~(\ref{geomsplit}) removes possible degeneracies that could be responsible for a reduced number of Lyapunov exponents. One-dimensional systems do not pose a problem for the intrinsic formalism, as we have already seen in the example of the inverted harmonic oscillator. 

%%%%%%%%%%%%%%%%%%%%%%%%%%%%%%%%%%%
\section{Conclusion}\label{sec_conclusion}
The use of geometric language clearly specifies the intrinsic ingredients of a given physical system, and at the same time clarifies which mathematical structures may be meaningfully used. In this article we have analyzed the stability of solutions of dynamical systems in geometric language. We have restricted ourselves to the important class of continuously evolving systems which can be formulated as vector flows $X$ over some smooth manifold $S$. Although the solutions simply are integral curves of the vector field $X$, this is not a restriction to first order systems; the order rather depends on the choice of $(S,X)$. Linear perturbations of reference solutions are characterized by vector fields~$\xi$ over $S$ that commute with $X$. Within this basic framework we have reconsidered the question of the stability of solutions under linear perturbations. 

For the analysis of global stability exists the extensive mathematical theory of Lyapunov exponents. Here the use of the geometric formulation nicely clarifies that global stability in general is not an intrinsic notion. It requires outside information that has to be added to the dynamical system~$(S,X)$. More precisely, a family of seminorms has to be given on the tangent bundle $TS$. This arbitrariness of choice is not sufficiently appreciated in the literature that applies the global stability concept.

In order to analyze the local stability of solutions of $(S,X)$ we have proposed a very intuitive new criterion that requires the additional choice of a linear connection on $TS$ to measure local accelerations. Hence also  local stability in general is not intrinsic to the dynamical system. The connection criterion for local stability is applicable to any dynamical system. For the case of second order systems we have shown how it reduces to the stability theory of Kosambi, Cartan and Chern. In fact, the new criterion is precisely equivalent to KCC theory for natural Lagrangian systems. 

Some dynamical systems contain sufficient geometrical structure to allow for completely intrinsic definitions of global and local stability. It is well-known that this is the case for geodesics on a Riemannian manifold. But notably, we have shown that this is also the case for general Lagrangian systems $(M,L)$. Writing them as dynamical systems $(S=TM,X)$ shows that the definition of global Lyapunov stability proceeds from the generalized Lagrange metric which simply is extended to the bundle~$TS$. Local stability according to the newly proposed connection criterion uses the Berwald type connection on $TS$ associated to the Lagrangian semispray $X$. No previous notion of local stability seems to cover general Lagrangian systems.
 
For natural Lagrangian systems given by Lagrangians quadratic in the velocities, the Jacobi-Maupertuis principle is commonly applied to translate solutions of fixed energy into geodesics of an associated Riemannian manifold. We have compared the resulting notions of geodesic stability with the intrinsic notions derived from the Lagrangian system. We have pointed out that geodesic stability suffers from various illnesses: boundaries of the Riemannian manifold that are curvature singular; the non-ability to resolve perturbations that modify energy; and the complete breakdown of the analysis for one-dimensional systems. None of these issues troubles the intrinsic definitions of stability in the Lagrangian picture.

The main achievement of this article lies in clarifying in a geometric framework the structures that underlie any discussion of the global or local stability of solutions of dynamical systems. It has emerged that the geometry of dynamical systems forces us to define a new criterion for local stability in terms of a choice of linear connection. Our comparison of this criterion to the geodesic stability of natural Lagrangian systems already shows a number of advantages of the intrinsic geometric approach. We expect that the practical relevance of our construction will be proven in future applications.

%%%%%%%%%%%%%%%%%%%%%%%%%%%%%%%%%%%
\acknowledgments
The authors wish to thank Christian Gross and Paul Smyth for stimulating discussions. They gratefully acknowledge full financial support from the German Research Foundation DFG through the Emmy Noether grant WO 1447/1-1.

%%%%%%%%%%%%%%%%%%%%%%%%%%%%%%%%%%%

%%%%%%%%%%%%%%%%%%%%%%%%%%%%%%%%%%%

\begin{thebibliography}{00}
\newpage

%\cite{Oseledec:1968} 
\bibitem{Oseledec:1968} 
  V.~I.~Oseledec, Trans.\ Mosc.\ Math.\ Soc. {\bf 19} (1968) 197.
  %% NOT IN SPIRES %%

%\cite{Pesin:1977}
\bibitem{Pesin:1977}
  Y.~B.~Pesin, 
  %``Characteristic Lyapunov Exponents and Smooth Ergodic Theory'', 
  Russian\ Math.\ Surveys {\bf 32} (1977) 55.
  %% NOT IN SPIRES %%

%\cite{Benettin:1976} 
\bibitem{Benettin:1976}
  G.~Benettin, L.~Galgani and J.~M.~Strelcyn, 
  %``Kolmogorov entropy and numerical experiments'', 
  Phys.\ Rev.\ A {\bf 14} (1976) 2338.
  %% NOT IN SPIRES %%
 
%\cite{Benettin:1980a} 
\bibitem{Benettin:1980a}
  G.~Benettin, L.~Galgani, A.~Giorgilli and J.~M.~Strelcyn, 
  Meccanica {\bf 15} (1980) 9.
  %% NOT IN SPIRES %% 
  
%\cite{Benettin:1980b} 
\bibitem{Benettin:1980b}
  G.~Benettin, L.~Galgani, A.~Giorgilli and J.~M.~Strelcyn, 
  Meccanica {\bf 15} (1980) 21.
  %% NOT IN SPIRES %% 

%\cite{Gozzi:1993tm}
\bibitem{Gozzi:1993tm}
  E.~Gozzi and M.~Reuter,
  %``Lyapunov exponents, path integrals and forms,''
  Chaos Solitons Fractals {\bf 4} (1994) 1117.
  %%CITATION = CSFOE,4,1117;%%   

%\cite{Caiani:1997yv}
\bibitem{Caiani:1997yv}
  L.~Caiani, L.~Casetti, C.~Clementi and M.~Pettini,
  %``Geometry of dynamics, Lyapunov exponents and phase transitions,''
  Phys.\ Rev.\ Lett.\  {\bf 79} (1997) 4361
  [arXiv:chao-dyn/9702011].
  %%CITATION = PRLTA,79,4361;%%

%\cite{Casetti:2000gd}
\bibitem{Casetti:2000gd}
  L.~Casetti, M.~Pettini and E.~G.~D.~Cohen,
  %``Geometric approach to Hamiltonian dynamics and statistical mechanics,''
  Phys.\ Rept.\  {\bf 337} (2000) 238.
  %%CITATION = PRPLC,337,238;%%

%\cite{Terzic:2002uj}
\bibitem{Terzic:2002uj}
  B.~Terzic and H.~E.~Kandrup,
  %``Semi-Analytic Estimates of Lyapunov Exponents in Lower-Dimensional
  %Systems,''
  Phys.\ Lett.\  A {\bf 311} (2003) 165
  [arXiv:astro-ph/0211248].
  %%CITATION = PHLTA,A311,165;%%

%\cite{Cerruti:1996}
\bibitem{Cerruti:1996}
  M.~Cerruti-Sola and M.~Pettini,
  %``Geometric description of chaos in two-degrees-of-freedom Hamiltonian systems'',
  Phys.\ Rev.\ E {\bf 53} (1996) 179. 
  %% NOT IN SPIRES %%

%\cite{Bari:1997zz}
\bibitem{Bari:1997zz}
  M.~Di Bari, D.~Boccaletti, P.~Cipriani and G.~Pucacco,
  %``Dynamical behavior of Lagrangian systems on Finsler manifolds,''
  Phys.\ Rev.\  E {\bf 55} (1997) 6448.
  %%CITATION = PHRVA,E55,6448;%%

%\cite{Cipriani:1998zz}
\bibitem{Cipriani:1998zz}
  P.~Cipriani and M.~Di Bari,
  %``Finsler Geometric Local Indicator of Chaos for Single Orbits in the
  %Henon-Heiles Hamiltonian,''
  Phys.\ Rev.\ Lett.\  {\bf 81} (1998) 5532.
  %%CITATION = PRLTA,81,5532;%%

%\cite{CuervoReyes}
\bibitem{CuervoReyes}
  E.~Cuervo-Reyes,
  %``General disagreement between the Geometrical Description of Dynamical 
  %In-stability -using non affine parameterizations- and traditional 
  %Tangent Dynamics'',
  arXiv:0807.1156 [math-ph].
  %% NOT IN SPIRES %%

%\cite{Cerruti:2008}
\bibitem{Cerruti:2008}
  M.~Cerruti-Sola, G.~Ciraolo, R.~Franzosi and M.~Pettini,
  %``Riemannian geometry of Hamiltonian chaos: hints for a general theory'',
  Phys.\ Rev.\ E {\bf 78} (2008) 046205.
  %% NOT IN SPIRES %%

\bibitem{Kosambi}
  D.~D.~Kosambi,
  %``Parallelism and path-spaces'',
  Math.\ Z. {\bf 37} (1933) 608.  
  %%NOT IN SPIRES%%

\bibitem{Cartan}
  E.~Cartan,
  %``Observations sur le m\'emoire pr\'ec\'edent'',
  Math.\ Z. {\bf 37} (1933) 619.
  %%NOT IN SPIRES%%

\bibitem{Chern}
  S.~S.~Chern,
  %``Sur la g\'eom\'etrie d'un syst\`eme d'\'equations diff\'erentielles du second ordre'',
  Bull.\ Sci.\ Math. {\bf 63} (1939) 206.
  %%NOT IN SPIRES%%

%\cite{Bucataru:2007b}
\bibitem{Bucataru:2007b}
  I.~Bucataru and R.~Miron,
  %``The geometry of systems of third order differential equations 
  %induced by second order Lagrangians'',
  arXiv:0705.3689 [math.DG].
  %% NOT IN SPIRES %%
  
%\cite{Harko:2008ak}
\bibitem{Harko:2008ak}
  T.~Harko and V.~S.~Sabau,
  %``Jacobi stability of the vacuum in the static spherically symmetric brane
  %world models,''
  Phys.\ Rev.\  D {\bf 77} (2008) 104009
  [arXiv:0803.4369 [gr-qc]].
  %%CITATION = PHRVA,D77,104009;%%
  
%\cite{Bucataru:1999}  
\bibitem{Bucataru:1999}
  I.~Bucataru,
  %``The Jacobi fields for a spray on the tangent bundle,''
  Novi\ Sad\ J.\ Math. {\bf 29} (1999) 69.
  %%NOT IN SPIRES%% 
  
%\cite{Yanobook}
\bibitem{Yanobook}
  K.~Yano and S.~Ishihara,
  {\it Tangent and cotangent bundles},
  Marcel Dekker, New York 1973.  
  %%NOT IN SPIRES%%
  
%\cite{Crampin}
\bibitem{Crampin}
  M.~Crampin,
  %``Tangent bundle geometry for Lagrangian dynamics'',
  J.\ Phys.\ A {\bf 16} (1983) 3755.
  %%NOT IN SPIRES%%
  
%\cite{Bucataru:2007a}
\bibitem{Bucataru:2007a}
  I.~Bucataru,
  %``Metric nonlinear connections'',
  Diff.\ Geom.\ App. {\bf 25} (2007) 335.
  %%NOT IN SPIRES%%
  
%\cite{Townsend:2004zp}
\bibitem{Townsend:2004zp}
  P.~K.~Townsend and M.~N.~R.~Wohlfarth,
  %``Cosmology as geodesic motion,''
  Class.\ Quant.\ Grav.\  {\bf 21} (2004) 5375
  [arXiv:hep-th/0404241].
  %%CITATION = CQGRD,21,5375;%%
  
%\cite{Elias:2007uy}
\bibitem{Elias:2007uy}
  L.~A.~Elias and A.~Saa,
  %``Homogeneous cosmologies and the Maupertuis-Jacobi principle,''
  Phys.\ Rev.\  D {\bf 75} (2007) 107301
  [arXiv:gr-qc/0702058].
  %%CITATION = PHRVA,D75,107301;%%
  
%\cite{Hrycyna:2005qp}
\bibitem{Hrycyna:2005qp}
  O.~Hrycyna and M.~Szydlowski,
  %``Different faces of chaos in FRW models with scalar fields: Geometrical
  %point of view,''
  Chaos Solitons Fractals {\bf 28} (2006) 1252
  [arXiv:gr-qc/0505155].
  %%CITATION = CSFOE,28,1252;%%  
  
\end{thebibliography}
\end{document}